\documentclass[lettersize,journal]{IEEEtran}
\usepackage{amsmath,amsfonts}
\usepackage{algorithmic}
\usepackage{algorithm}
\usepackage{array}
\usepackage[caption=false,font=normalsize,labelfont=sf,textfont=sf]{subfig}
\usepackage{textcomp}
\usepackage{stfloats}
\usepackage{url}
\usepackage{verbatim}
\usepackage{graphicx}
\usepackage{overpic}
\usepackage{cite}

\begin{document}

\title{Snapshot High Dynamic Range Imaging with a Polarization Camera}

\author{Mingyang Xie*, Matthew Chan*, Christopher Metzler
\thanks{* Equal contribution}%
\thanks{
The authors are with the Department of Computer Science, University of Maryland, College Park, MD, 20742 (e-mail: mingyang@umd.edu; mattchan@umd.edu; metzler@umd.edu).}}

\maketitle

\begin{abstract}
High dynamic range (HDR) images are important for a range of tasks, from navigation to consumer photography. Accordingly, a host of specialized HDR sensors have been developed, the most successful of which are based on capturing variable per-pixel exposures. In essence, these methods capture an entire exposure bracket sequence at once in a single shot. This paper presents a straightforward but highly effective approach for turning an off-the-shelf polarization camera into a high-performance HDR camera. By placing a linear polarizer in front of the polarization camera, we are able to simultaneously capture four images with varied exposures, which are determined by the orientation of the polarizer. We develop an outlier-robust and self-calibrating algorithm to reconstruct an HDR image (at a single polarity) from these measurements. Finally, we demonstrate the efficacy of our approach with extensive real-world experiments.
\end{abstract}


\section{Introduction}\label{sec:introduction}

\IEEEPARstart{T}{he} dynamic range of digital cameras is constrained to a limited range---far narrower than the human eye. 
As a result, when everyday scenes are captured by a standard low-dynamic-range (LDR) camera, a large number of pixels will be over-exposed or under-exposed and information will be lost.
The most common approach to record HDR scenes is to capture multiple images with varying exposures~\cite{Picard95onbeing,Debevec:1997,mertens2009exposure}.
However, such multi-shot HDR techniques are susceptible to motion-induced ghosting artifacts.

Over the last two decades, a plethora of single-shot HDR imaging techniques have been developed to avoid this problem. 
One can optically split an image so as to capture multiple exposures on multiple sensors simultaneously~\cite{McGuire:2007, Tocci:2011},
one can perform glare-encoding to recover bright highlights~\cite{Rouf:2011,metzler2020deep,sun2020learning}, one can use exotic modulo sensors and solve a phase unwrapping problem to recover an HDR scene~\cite{Zhao:2015, Zhou2020UnModNetLT, So2021MantissaCamLS}, and one can also simply capture an over-exposed LDR image and then use a CNN to hallucinate the missing details~\cite{HDRCNN}.      

Arguably the most successful approach to date for single-shot HDR imaging is to apply per-pixel exposures~\cite{nayar2000high}.  
This approach is simpler and more robust than optical encoding based methods and, unlike modulo sensors, does not suffer from any brightness ambiguities at edges/discontinuities in an image. 

Sony recently commercialized this technology and now sells dedicated HDR image sensors, such as the IMX490. In this work we demonstrate how one can achieve similar functionality using an off-the-shelf polarization camera.

\begin{figure}[!t]
\centering
\includegraphics[width=\columnwidth]{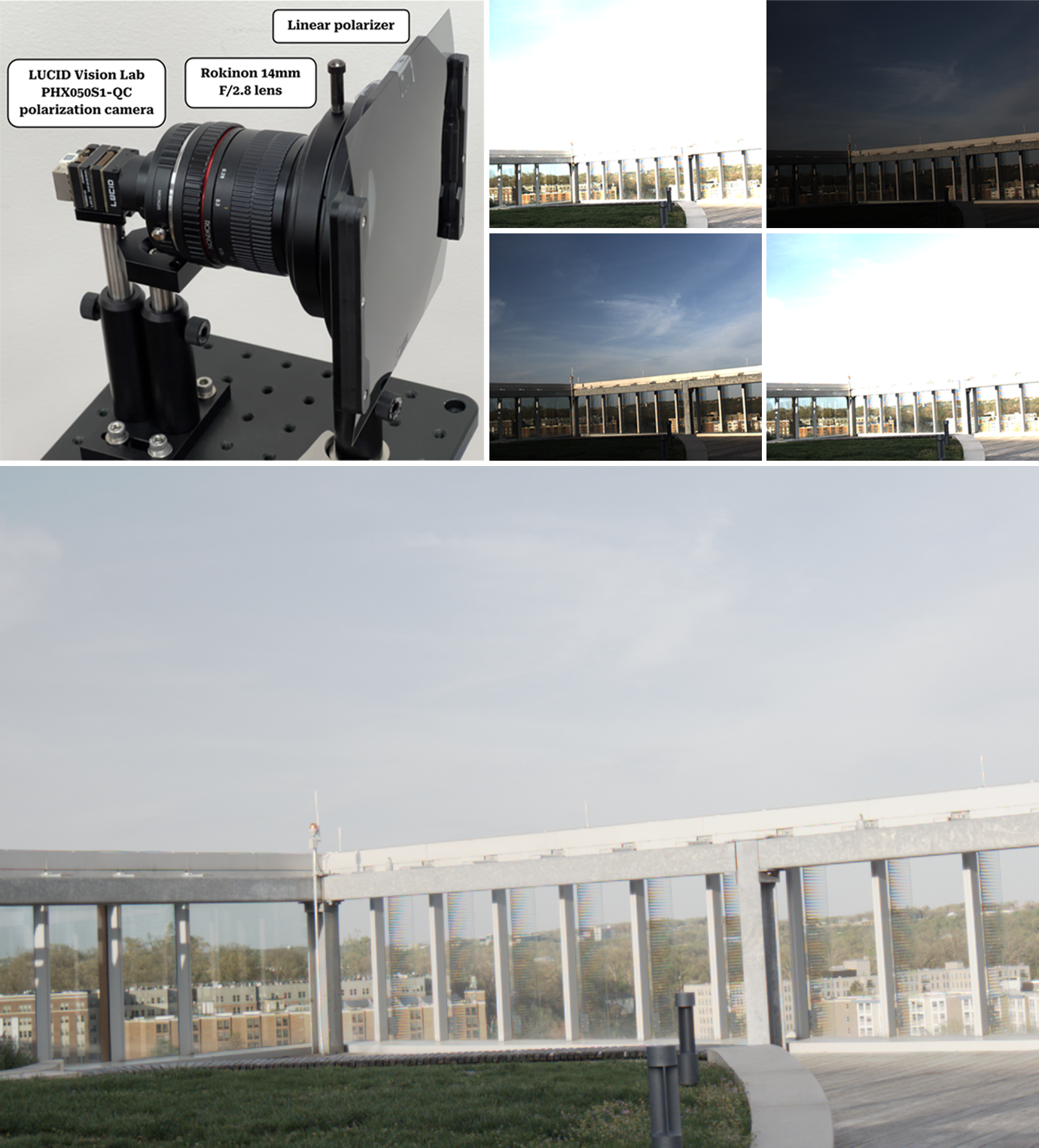}
\caption{\textbf{Single-shot HDR reconstruction using a polarization camera.} Our method combines a polarization camera and linear polarizing filter (top left) to capture 4 multi-exposure images simultaneously (top right) and reconstruct HDR images (bottom).}
\label{fig:teaser}
\end{figure}

\subsection{Our Contribution}
We demonstrate how an off-the-shelf polarization camera can be (reversibly) repurposed to perform snapshot HDR imaging. Conceptually, our hardware setup works similarly to Sony's HDR sensors---albeit with an adjustable dynamic range. 

This functionality is achieved by placing a linear polarizer in front of the camera's lens (see Figure~\ref{fig:teaser}). Combining this polarizer with the polarization camera's Bayer-like polarization grid, we effectively capture four exposures---one for each of the four polarizer angles ($0^\circ, 45^\circ, 90^\circ, 135^\circ$)---in a single-shot. Under this construction, the residual between angle of the linear polarizer and the angle of the polarization array determines each image's relative exposure~\cite{BENNETT2005190}. Images captured by approximately orthogonal polarization angles (e.g. when the linear polarizer is rotated to $90^\circ$ and we read out the $0^\circ$ polarized image) result in images with an extremely low exposure (see Figure~\ref{fig:malus}). 

This paper makes the following contributions:
\begin{itemize}
    \item We develop a novel methodology for capturing high dynamic range information with minimal hardware modification to an off-the-shelf polarization camera.
    \item We introduce an HDR reconstruction algorithm that is robust to outliers and light leakage caused by imperfect polarizers.
    \item We analyze the effects that imperfect polarizers and highly polarized scenes have on our reconstructions.
    \item We conduct extensive experiments demonstrating that our method significantly outperforms state-of-the-art software-only HDR reconstruction algorithms, as well as existing polarization-camera based HDR imaging techniques~\cite{zhou2023polarization}.
\end{itemize}

\section{Related Work}

\subsection{Multi-frame HDR Reconstruction}

Exposure bracketing based HDR reconstruction methods~\cite{debevec2008recovering,mertens2007exposure,yan2019multi} fuse well-exposed pixels from images captured at different exposure stops to produce an HDR image.

While effective for static scenes, these techniques are susceptible to ghosting artifacts caused by motion blur and camera shake. Imaging in low-light further exacerbates this issue as the shutter must remain open for longer intervals of time to properly expose pixels.

As part of their work on the Google Pixel, Hasinoff et al.~proposed HDR+~\cite{HDRplus}, which relies on a burst of short-exposure images instead of a traditional exposure bracket. While under-exposed regions of these images are influenced by photon shot noise, this effect is mitigated by averaging intensities across captures. The primary benefit of this method is that it's inherently less affected by ghosting artifacts since the shutter is open for a shorter duration. Still, the HDR ratio achievable by this method is capped due to the limited dynamic range of captured exposure stops.

Removing the need for exposure calibration altogether, Exposure Fusion~\cite{mertens2009exposure} fuses LDR images according to heuristics such as contrast, saturation, and well-exposedness. We note, however, that while this method produces images with an HDR-like appearance, the resulting images do not accurately reflect the scene's true intensity.

\subsection{Software-based Snapshot HDR Reconstruction}
Recovering an HDR image from a single LDR image is an under-constrained problem. Over-exposed pixels contain very little information and under-exposed pixels are overwhelmed by noise. Deep learning based methods~\cite{HDRCNN,chen2021hdrunet} address this problem by using learned priors. However, while these methods excel at hallucinating plausible information in poorly exposed regions, they perform poorly on out-of-distribution data.

\subsection{Hardware-based Snapshot HDR Reconstruction}
Addressing the challenge of snapshot HDR reconstruction from a different perspective, some methods leverage exotic imaging hardware to encode HDR information form a single capture. Nayar et al.~\cite{nayar2000high} proposed a split-pixel sensor design capable of capturing four different stops of exposure simultaneously. So et al.~\cite{So2021MantissaCamLS} and Zhao et al.~\cite{zhao2015unbounded} prevent clipping by capturing HDR scenes with a modulo camera and phase unwrapping the result. Typically, methods of this nature generalize well across scenes, but there are some caveats. For instance, So et al.'s MantissaCam struggles to recover regions with sharp discontinuities, such as windows. 

More recently, Zhou et al. used a polarization camera to perform HDR imaging~\cite{zhou2023polarization}. Unlike our method, their approach does not use an additional polarizing filter in front of the lens. Even without an added filter, each of the four polarizers in the camera attenuates incoming light in a slightly different way. Using this intensity information and leveraging polarity information, Zhou et al. train their Pol-HDR network to recover HDR images from a snapshots of four polarized LDR images. The lack of an additional polarizing filter drastically reduces the range of exposures covered, resulting in lower dynamic range reconstructions compared to our method.

\begin{figure}[t]
    \centering
        \includegraphics[width=0.8\columnwidth]{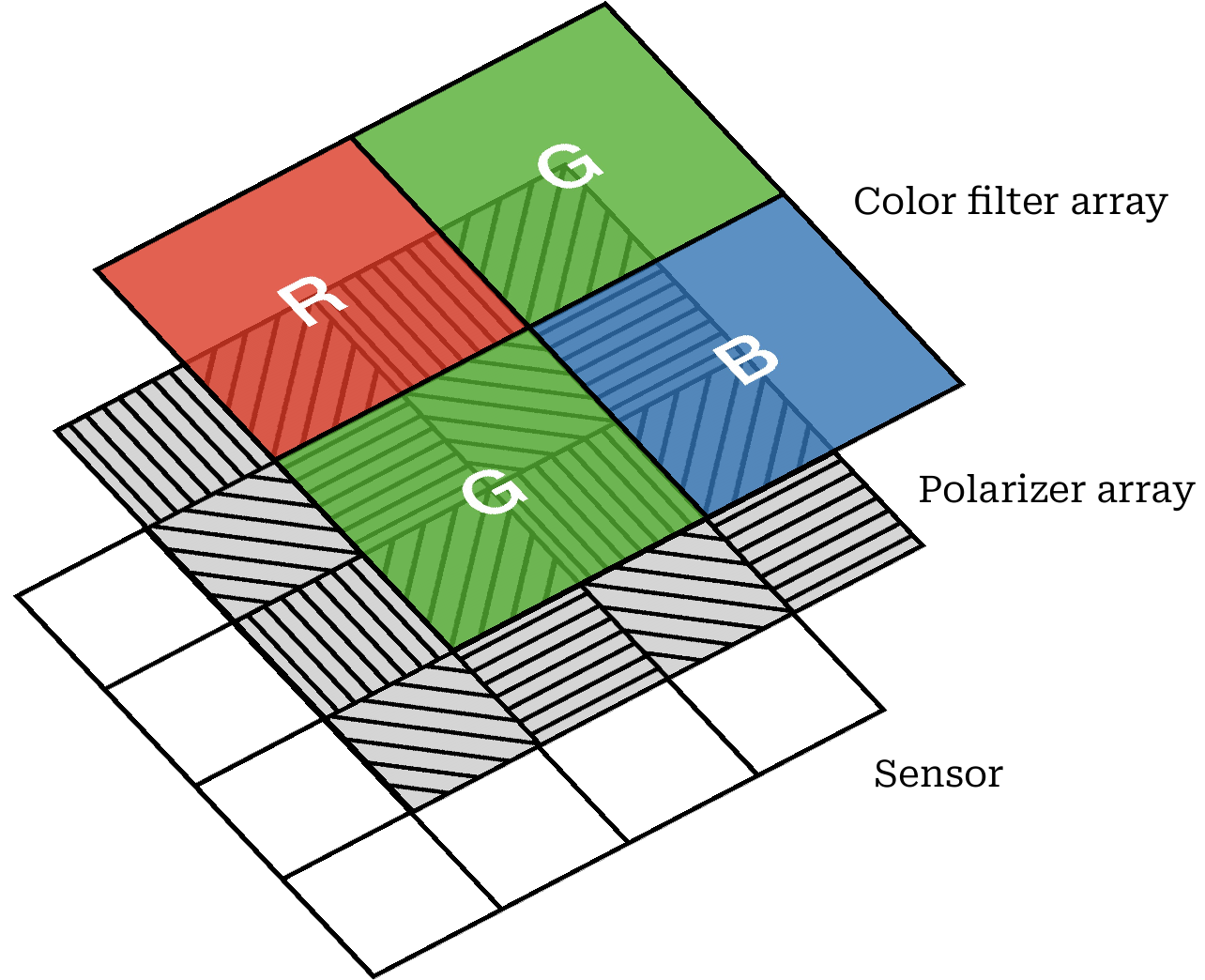}
        \caption{\textbf{Polarization and color encoding in polarization cameras.} Polarization cameras contain a stacked filter grid to encode information about both polarity and color in each pixel. The top-level RGGB Bayer grid encodes color information while the underlying polarizer grid encodes polarity.}
    \label{fig:PolarizationCamera}
\end{figure}

\section{Proposed Method}

\subsection{Hardware}
\label{sec:hardware}

Polarizers filter incident light such that light waves of a specific polarity are transmitted while orthogonally polarized light waves are obstructed. 
Polarization cameras use arrays of polarization filters and color filters to encode the polarity and color of each capture pixel. 

Specifically, each $4\times4$ grid of pixels is composed of four $2\times2$ pixel blocks, creating an RGGB Bayer pattern. Each pixel in these $2\times2$ blocks contains its own Bayer-like pattern of polarizing filters, arranged in 
$\{\frac{\pi}{2}, \frac{\pi}{4}, \frac{3\pi}{4}, 0 \}$ order. 
See Figure~\ref{fig:PolarizationCamera}.%

\begin{figure}[t]
    \centering
    \vspace{-8pt}
    \includegraphics[width=0.8\columnwidth]{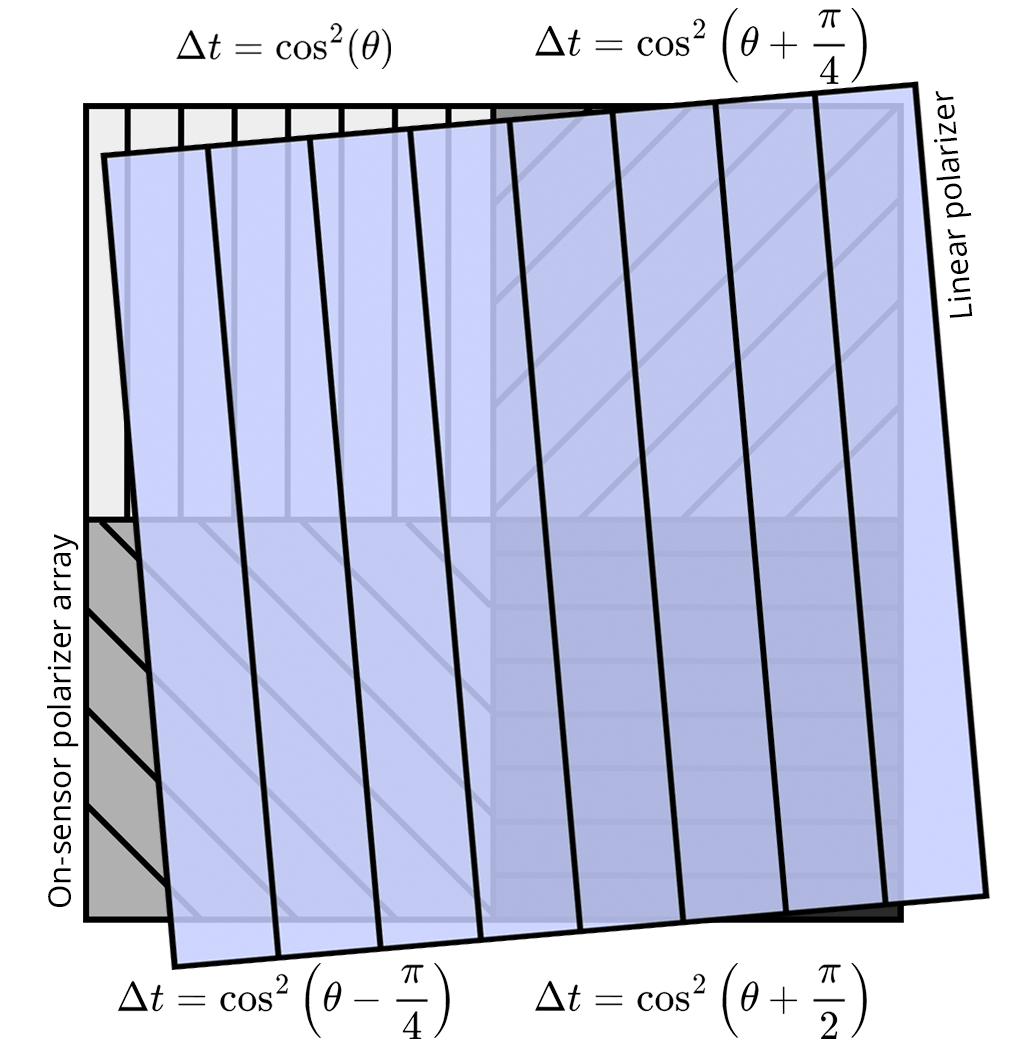}
    \caption{\textbf{Malus' law.} The more that elements of the on-sensor polarizer array are aligned with the linear polarizer, the more light they transmit. Likewise, the more orthogonal they become, the more they attenuate incident light. Transmission is proportional to cosine squared of the angle $\theta$ between the polarizers.}
    \label{fig:malus}
\end{figure}

In this work, we place an additional linear polarizing filter in front of the lens of a polarization camera to modulate the intensity of light transmitted onto the sensor. According to Malus' law, the intensity of each of these pixels is attenuated according to the square of the cosine of the angle between the two overlapping polarizers~\cite{collett2005field}. This variable attenuation enables the simultaneous capture of four exposure stops---rather than polarities---in a single shot (see Figure~\ref{fig:malus}).

\subsection{Problem Statement}
Digital cameras measure the intensity of incident light $I_0$ by exposing the sensor for some time $\Delta t$ to obtain a raw image
\begin{align}
    I(\Delta t) &= f(I_0\Delta t)
\end{align}
where $f$ is the clipping operator $\text{max}(\text{min}(x, 255), 0)$ corresponding to the 8-bit depth of our hardware. 

Assuming that $\Delta t$ is known, reconstruction of an HDR image becomes a straightforward process---simply compute a weighted sum of LDR images based on exposure. The introduction of an additional polarizing filter (at an unknown orientation) in our optical system complicates this process, however, as the exposure of each image is no longer known. 

Although HDR reconstruction algorithms that do not require explicit knowledge of $\Delta t$ exist (e.g.,~\cite{mertens2007exposure}), these algorithms do not perform metric HDR but rather produce physically inaccurate---albeit visually pleasing---images. Thus, the question that we seek to answer becomes, ``how can we recover the relative exposures of the four images we capture?''

\subsection{Exposure Calibration}

With the addition of a polarizer in front of the lens, our polarization camera simultaneously captures a set of four images $S=\{I(\theta_1), I(\theta_2), I(\theta_3), I(\theta_4)\}$ corresponding to polarization angles $\{0, \frac{\pi}{4}, \frac{\pi}{2}, \frac{3\pi}{4}\}$ respectively. Each image has a transmission rate dependent on $\theta$ that determines their exposure $\Delta t = \text{cos}^2\theta_i$.

Assuming the camera’s raw measurements are accessible, we can ignore non-linearities in the polarization camera’s image processing pipeline (ISP) and %
express the intensities measured by each of the four pixel polarities as

\begin{align}
    \notag I(\theta_1) &= f(I_0\text{cos}^2(\theta_1))\\
    \notag I(\theta_2) &= f(I_0\text{cos}^2(\theta_2))\\
    \notag I(\theta_3) &= f(I_0\text{cos}^2(\theta_3))\\
    I(\theta_4) &= f(I_0\text{cos}^2(\theta_4)).
    \label{eqn:polarization}
\end{align}

We estimate the exposure of each image $I(\theta_i)$ by observing that pixels in the on-sensor polarizer array share the following relationship:

\begin{align}
    \theta_3 &= \theta_1 + \frac{\pi}{2}, \label{eq:theta_0_theta_2} \\
    \theta_4 &= \theta_2 + \frac{\pi}{2}.
\end{align}

For example, using Equation~\eqref{eq:theta_0_theta_2}, we can express images $I(\theta_1)$ and $I(\theta_3)$ in terms of $\theta_1$:
\begin{align}
    I(\theta_1) &= f(I_0\text{cos}^2\theta_1) \\
    I(\theta_3) &= f(I_0\text{sin}^2\theta_1)
\end{align}

For pixels for which both $I(\theta_1)$ and $I(\theta_3)$ are properly exposed, we obtain a closed-form solution for estimating angle $\theta_1$
\begin{equation}
    \hat{\theta}_1 = \text{arctan}\left(\sqrt{\frac{I(\theta_3)}{I(\theta_1)}}\right),
    \label{eq:closed-form}
\end{equation}

which allows us to estimate the exposure: $\text{cos}^2\hat{\theta}_1$. We estimate $\hat{\theta}_2, \hat{\theta}_3,$ and $\hat{\theta}_4$ using analogous expressions.

It is important to note that angle estimates $\hat{\theta}$ are computed over all $N$ pixels of the image. We aggregate these estimates (and reject under and over-exposed pixels) by taking their mode to obtain our final (scalar) estimate
\begin{equation}
    \hat{\theta}_{i} = \text{Mode}(\hat{\theta}_{i,j}), j \in [1, N].
\end{equation}

\begin{figure}[t]
    \centering
    \includegraphics[width=\columnwidth]{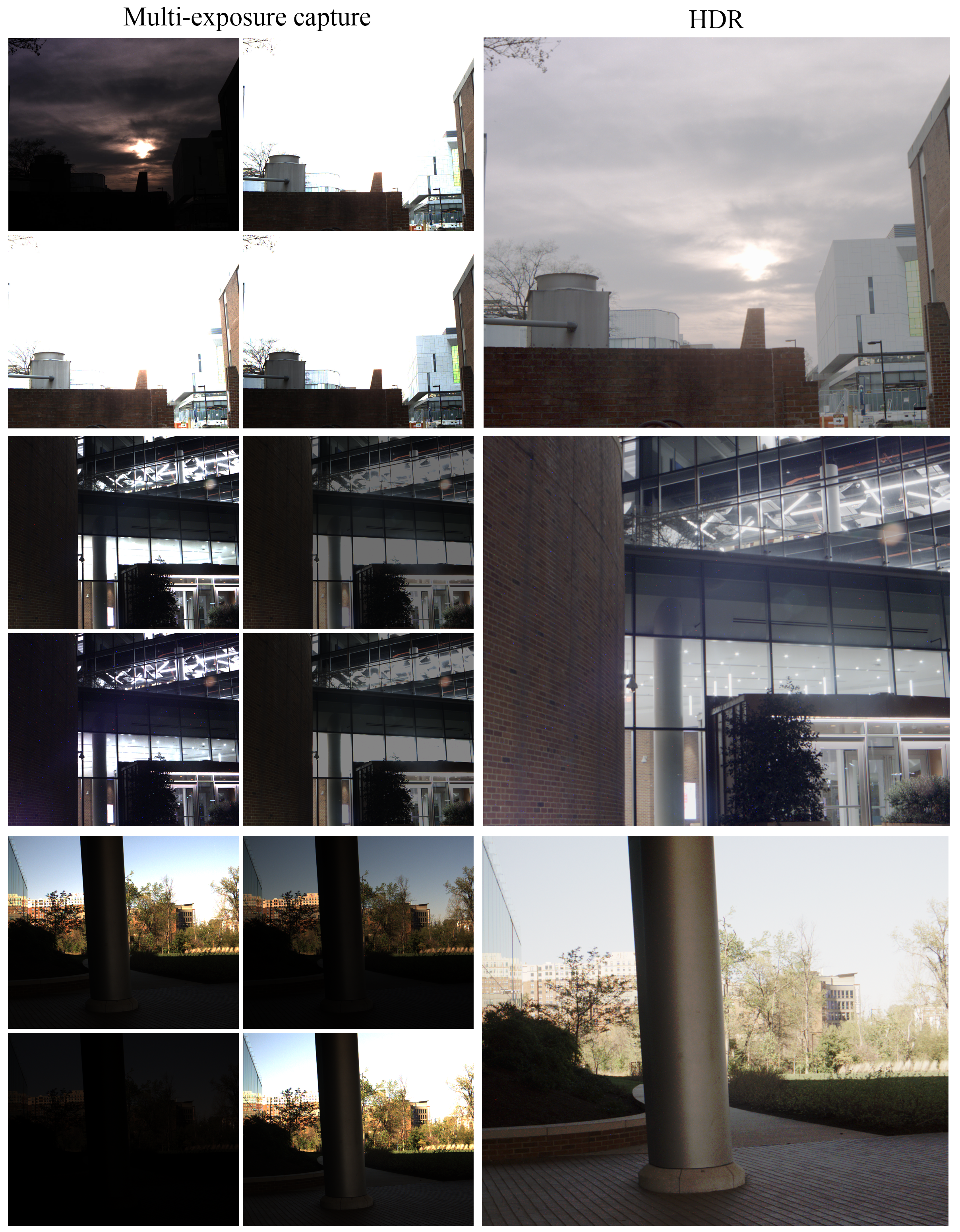}
    \caption{\textbf{Fusing LDR into HDR images.} Our polarization camera captures 4 multi-exposure images (left column). These are subsequently merged together to produce an HDR image (right column).}
    \label{fig:exposure-bracket}
\end{figure}

Having acquired these values, reconstruction of the HDR image boils down to the following steps: (1) scale each LDR image by its exposure, (2) mask out overexposed and under-exposed regions of the image, and (3) compute a normalized weighted sum over the masked and scaled LDR images:\\

Given a set of estimated angles $\hat{\theta}_i$ and corresponding LDR images, we first rescale the intensity of each LDR image $I(\theta_i)$ using
\begin{equation}
    I'(\theta_i) = \frac{I(\theta_i)}{\text{cos}^2\hat{\theta_i}}.
\end{equation}

We subsequently assign each pixel in the LDR image a corresponding weight indicating its well-exposedness. Specifically, pixels outside of the minimum and maximum intensity range $[p_{min}, p_{max}]$ are considered poorly exposed and receive zero weight according to the indicator function
\begin{equation}
    w(x) = \begin{cases}
        1, p_{min} \leq x \leq p_{max} \\
        0, \text{otherwise}
    \end{cases}
\end{equation}

Finally, we reconstruct the HDR result $H$ using a normalized weighted sum over the scaled images
\begin{equation}
    H = \frac{\sum_i^4 w(I'(\theta_i)) \odot I'(\theta_i)}{\sum_i^4 w(I'(\theta_i)) + \epsilon}
    \label{eq:classical-hdr}
\end{equation}
where $\odot$ is the Hadamard product and $\epsilon$ is a small positive value which avoids division by zero.

\begin{figure*}
    \centering
    \includegraphics[width=.93\textwidth]{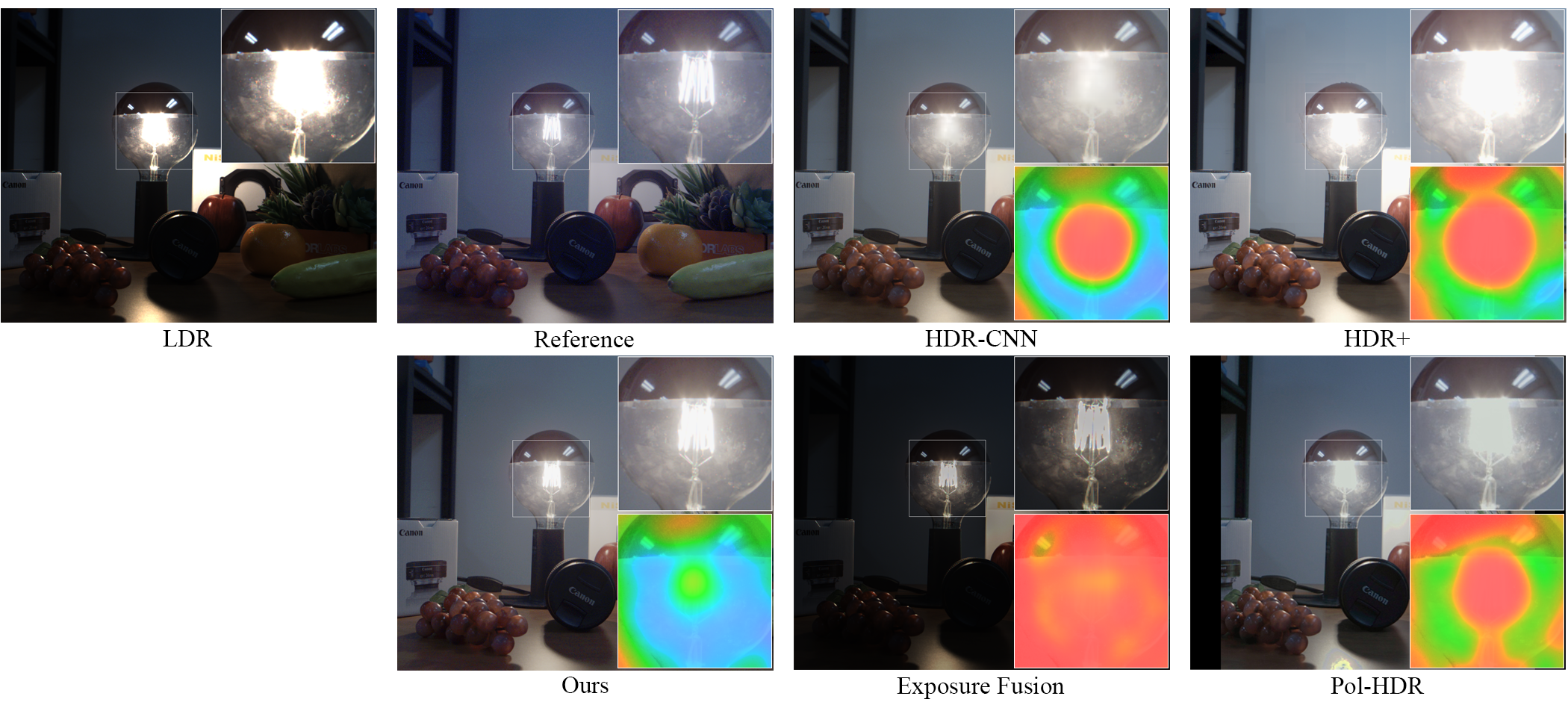}
    \includegraphics[width=.93\textwidth]{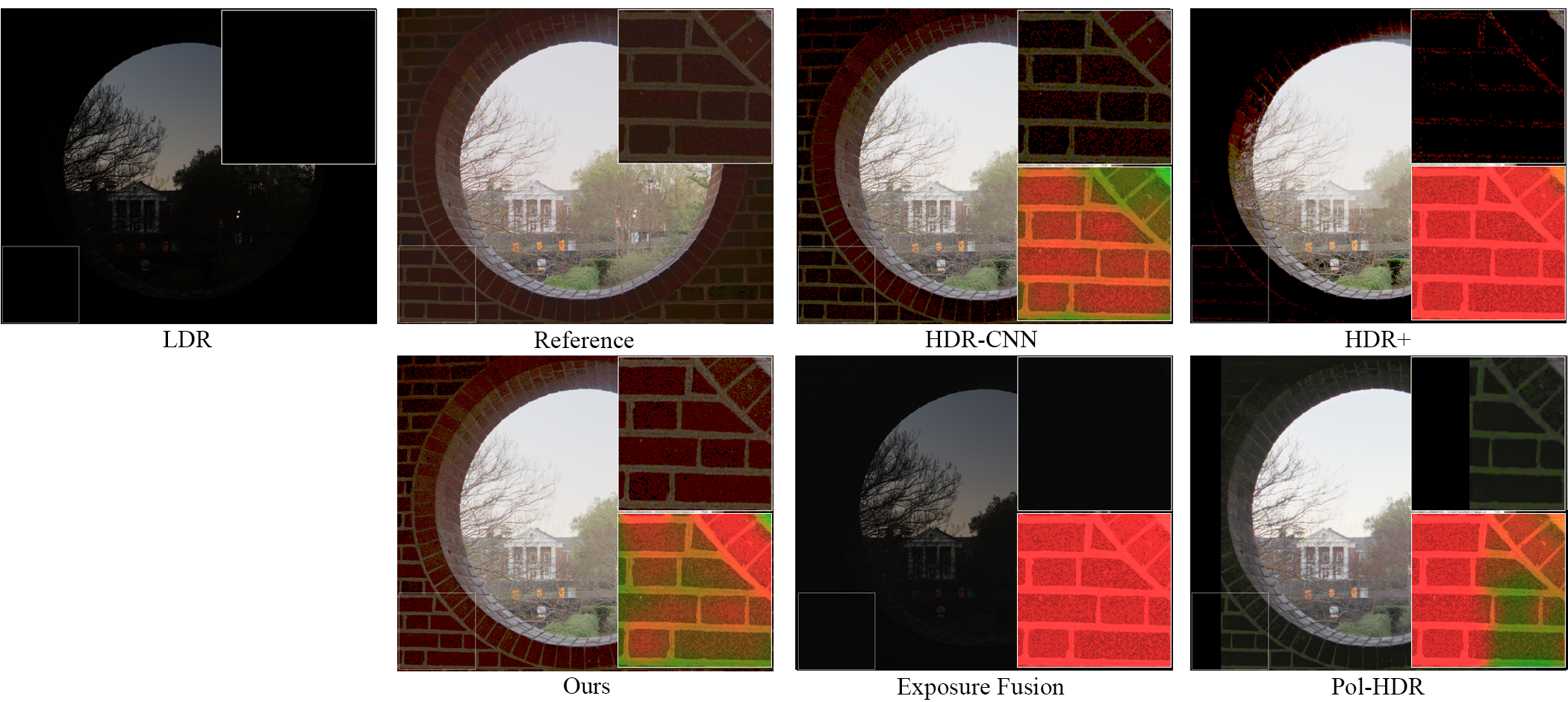}
    \includegraphics[width=.93\textwidth]{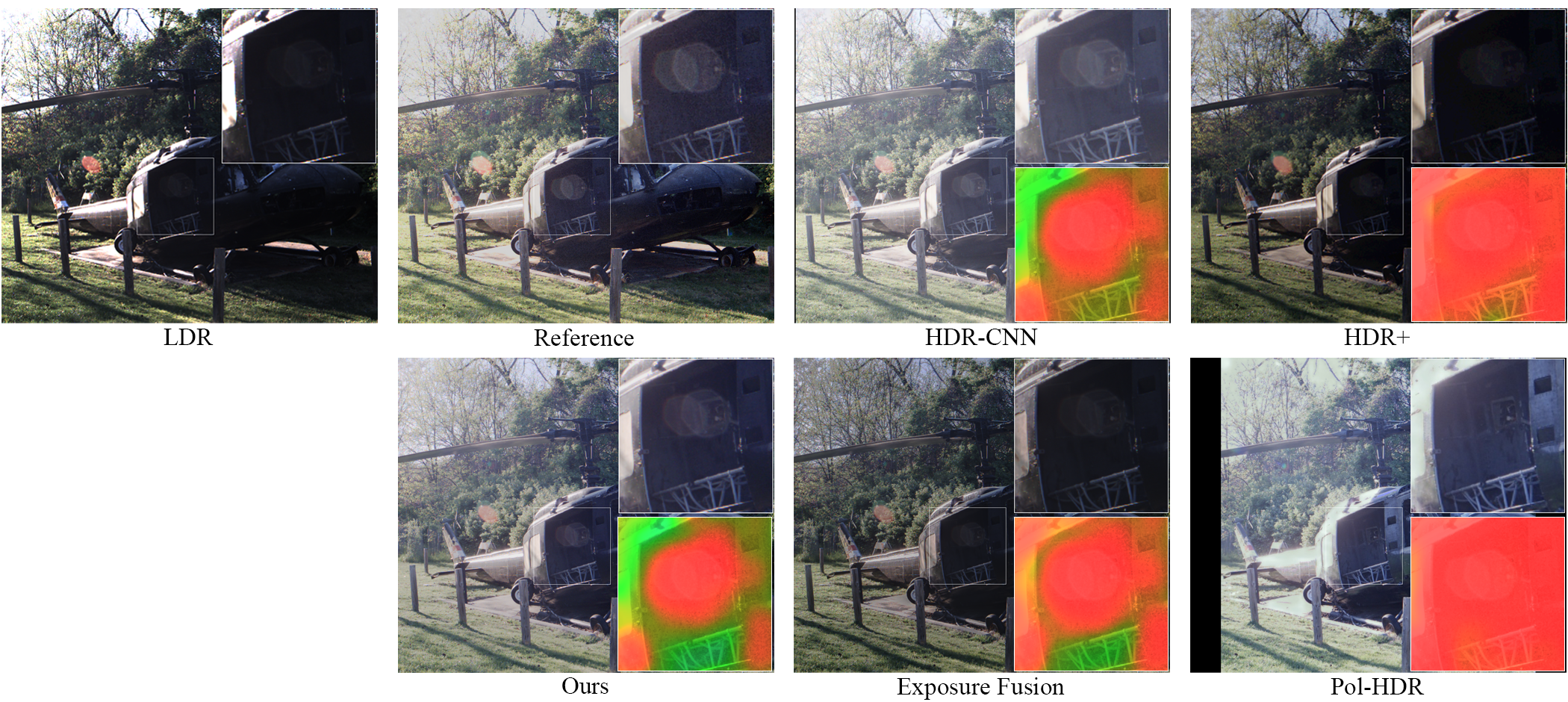}
    \caption{\textbf{Qualitative results.} We compare our method against state-of-the-art multi-shot (HDR+~\cite{HDRplus}), learning-based (HDR-CNN~\cite{HDRCNN}) and polarization-based (Pol-HDR~\cite{zhou2023polarization}) HDR reconstruction algorithms. Insets show zoomed in regions and $P_{det}$ heat maps~\cite{mantiuk2011hdr}. Red pixels indicate high probabilities of detecting differences against the ground truth. Blue pixels indicate low detection probabilities.}
    \label{results}
\end{figure*}

\begin{figure*}[t]
    \centering
    \begin{tabular}{|l||cccc|c|}
        \hline
         & HDR-CNN & HDR+ & Pol-HDR & Exposure Fusion & Ours \\
        \hline
        PSNR $\uparrow$ & 21.36 & 18.49 & 15.37 & 13.05 & \textbf{25.25} \\
        \hline
        SSIM $\uparrow$ & 0.85 & 0.69 & 0.65 & 0.60 & \textbf{0.88} \\
        \hline
        MS-SSIM $\uparrow$ & \textbf{0.90} & 0.83 & 0.73 & 0.77 & \textbf{0.90} \\
        \hline
        Q-score $\uparrow$ & 6.12 & 5.97 & 3.25 &  5.26 & \textbf{7.40} \\
        \hline
    \end{tabular}

    \caption{\textbf{Quantitative results.} Image quality scores of each baseline are averaged across all scenes in our dataset. Each metric is computed on the tone mapped HDR results. Best scores are shown in bold.}
    \label{fig:comparison-table}
\end{figure*}

\section{Experiments}

\subsection{Hardware}
Data in our experiments are captured on a LUCID Vision Lab PHX050S1-QC polarization camera with a \$38 off-the-shelf linear polarizer attached. All images are shot in RAW format at $1224\times1024$ resolution and each scene is imaged across 11 different exposure settings (covering over 8 stops of exposure). In total, we collect and label data from 12 natural scenes to form our HDR dataset.

\subsection{Setup and Baselines}
We evaluate our single-shot HDR method against both single-shot~\cite{zhou2023polarization, mertens2007exposure,HDRCNN} and multi-shot~\cite{HDRplus} approaches on a variety of scenes. In each scene, we calibrate the linear polarizer's orientation to span the broadest dynamic range possible and capture an exposure bracket $B=\{S_1, \cdots, S_{11}\}$ consisting of 11 snapshots using different exposure times. Each snapshot $S_k$ contains 4 images $I(\theta_i)$, one image per polarity.

Using Equation~\eqref{eq:classical-hdr}, we reconstruct the ground truth HDR image from LDR images $I(\theta_i)$ in $B$ belonging to a single polarity $\theta_i$. Then, we produce both our result and the Exposure Fusion~\cite{mertens2007exposure} baseline result using images $I(\theta_1), \cdots, I(\theta_4)$ from a single snapshot $S_k$. We also run HDR-CNN~\cite{HDRCNN} on individual images from $S_k$ and use the highest scoring reconstruction as a baseline. 

To generate our HDR+~\cite{HDRplus} baseline results, we capture a separate burst $B_{hdrplus}$ of 10 short exposure images. We also capture a separate snapshot $S_{polhdr}$ with the linear polarizer off to produce our Pol-HDR~\cite{zhou2023polarization} baseline results. All HDR reconstructions are tone mapped (where appropriate) using Reinhard's tone mapping operator~\cite{reinhard2010high} for better viewing on digital displays. For fair comparison, we apply the same tone mapping operator to HDR images for each scene.

Each method is benchmarked using peak signal-to-noise ratio (PSNR), structural similarity index (SSIM)~\cite{ssim}, multi-scale structural similarity index (MS-SSIM)~\cite{msssim}, Q-score, and $P_{map}$~\cite{mantiuk2011hdr}. For context, Q-score is an image quality indicator that behaves similarly to MS-SSIM. On the other hand, $P_{map}$ is a 2D heat map where the color of each pixel indicates the probability of a human observer detecting differences against the ground truth. Red pixels represent a high detection likelihood and blue pixels represent a low detection likelihood.

\subsection{Results}
Of all the methods, Exposure Fusion~\cite{mertens2007exposure} produces the lowest scoring results due to its reliance on heuristics like saturation and contrast to merge images. This causes the relative color and brightness of objects in the HDR results to consistently differ from the ground truth. Pol-HDR~\cite{zhou2023polarization} also scores relatively low on our benchmark, as it runs on images captured with the linear polarizer off, which severely limits the amount of intensity information available. Note that their method requires images with 1:1 aspect ratio as input, so we zero-pad their results before evaluation.

Google's HDR+~\cite{HDRplus} algorithm scores relatively well and produces reasonable looking HDR images. However, its results suffer from the limited dynamic range of the low exposure burst $B_{hdrplus}$. As a result, details in under-exposed regions are barely discernible (see the brick pattern in the circular window scene of Figure~\ref{results}).

\begin{figure}
    \centering
    \vspace{1cm}
    \begin{overpic}[width=0.24\textwidth]{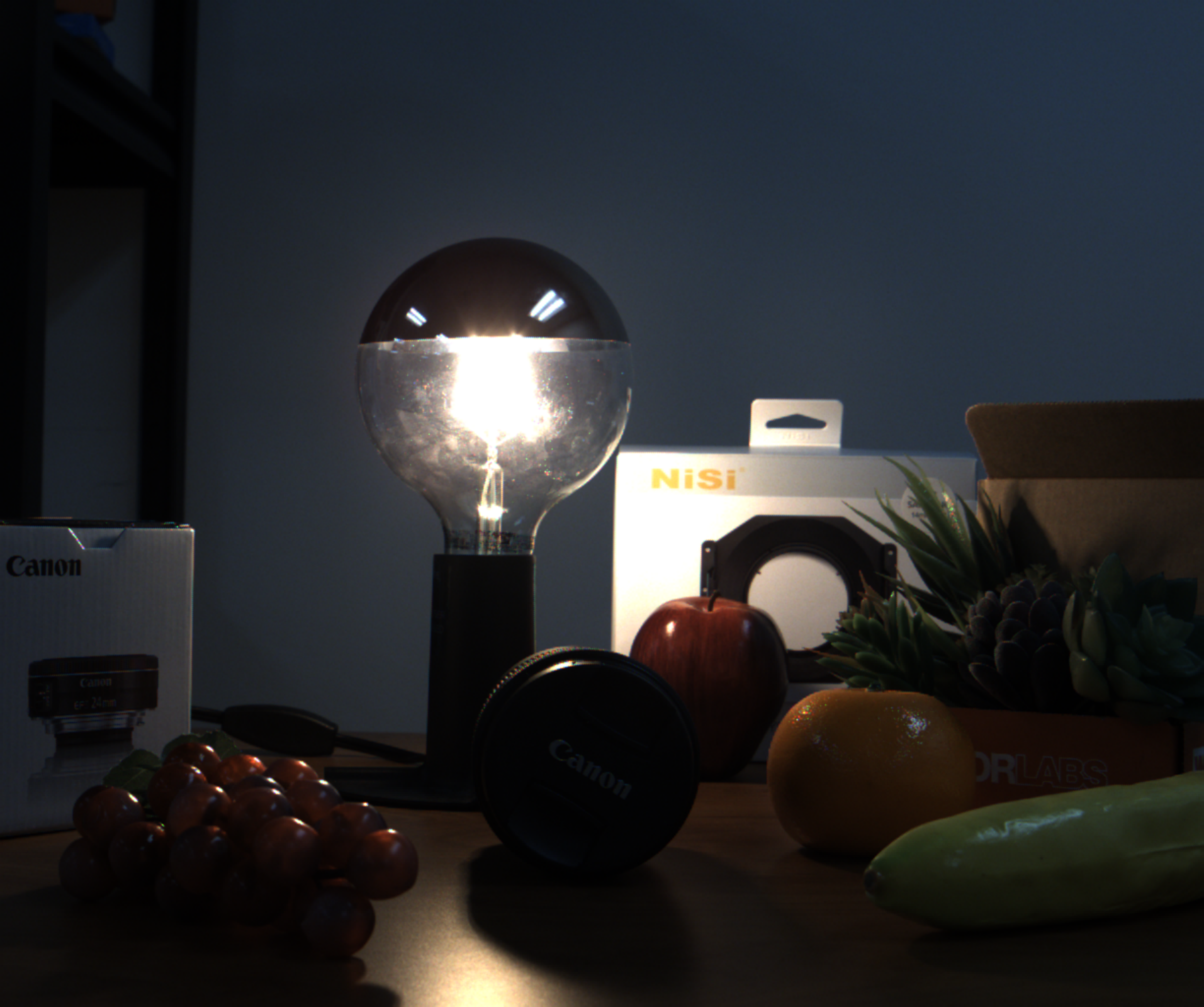}
		\put (18,87) {{\centering Parallel Polarizers}}
    \end{overpic}
    \begin{overpic}[width=0.24\textwidth]{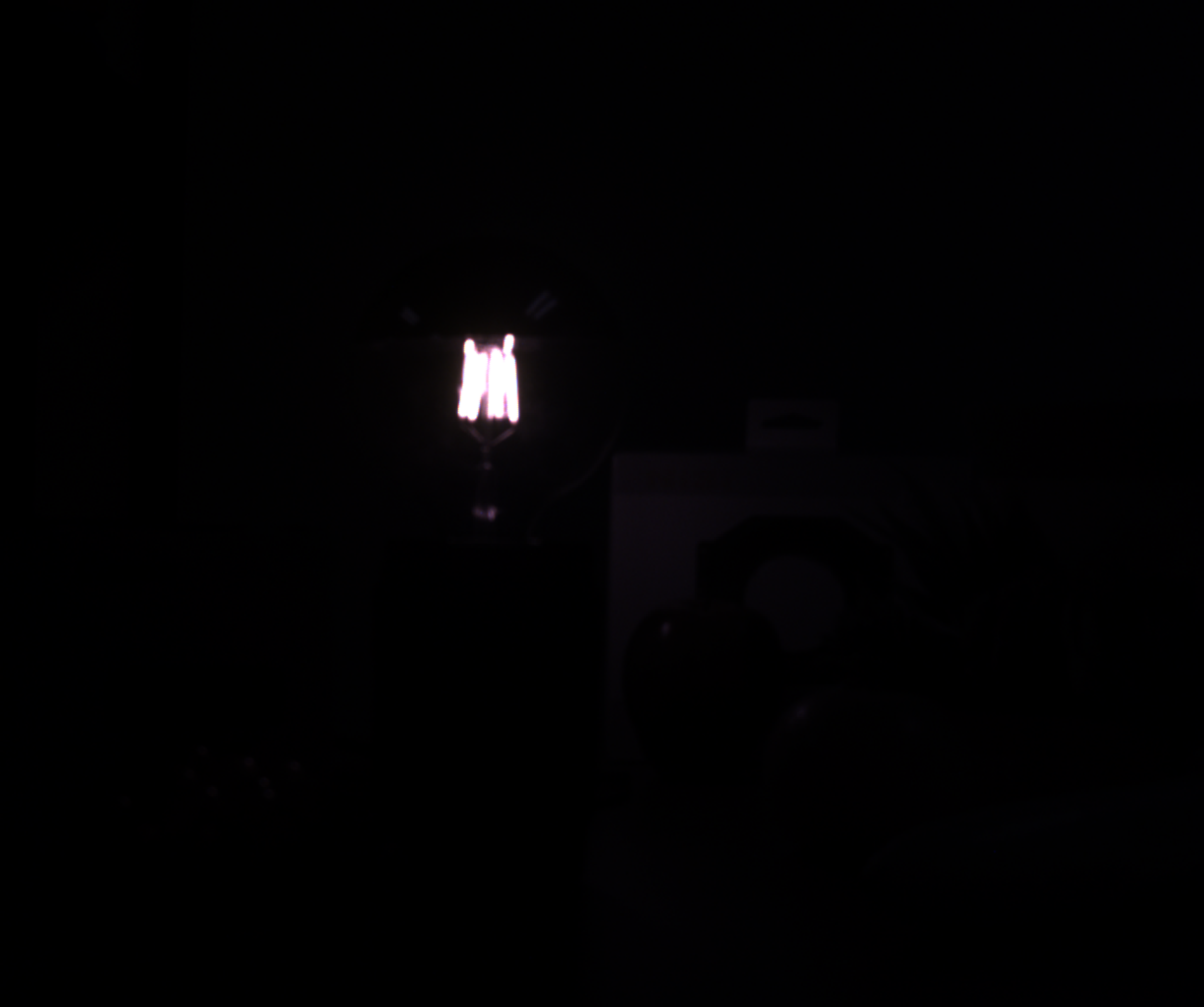}
		\put (15,87) {{\centering Orthogonal Polarizers}}
    \end{overpic}
    \caption{\textbf{Imperfect polarizers.} 
    In theory, polarizers perfectly polarize incoming light so that no light is transmitted across two orthogonal polarizers. However, imperfections occur during the manufacturing process such that polarizers are never $100\%$ efficient. We demonstrate this by imaging a light bulb (left) through orthogonal polarizers. The captured image (right) visibly contains non-zero pixels around bright regions of the scene.}
    \label{fig:light-leakage}
\end{figure}

Compared to other baselines, HDR-CNN~\cite{HDRCNN} was the highest performing method despite taking only a single frame $I(\theta)$ as input. It reasonably recovers information in under and over exposed regions of the image using learned priors. For example, in the table lamp scene, HDR-CNN suppresses over-exposed pixels around the incandescent bulb, making it appear more like a street lamp.

Our method scores the highest across all metrics (see Figure~\ref{fig:comparison-table}) and achieves a 18\% higher average PSNR than the next best method. Furthermore, $P_{map}$ scores indicate that our method produces images that qualitatively resemble the ground truth. For example, both the shape of the tungsten filament in the table lamp and the brick pattern around the circular window are well-exposed in our result.

\section{Analysis}

\subsection{Imperfect Polarizers}
In theory, angle estimates $\hat{\theta}$ computed using Equation~\eqref{eq:closed-form} recover the exact angle $\theta$ and exposure $\text{cos}^2\theta$ of each image. In practice, however, we observe errors in these estimates which stem from the imperfections present in real polarizers~\cite{huang2010attenuation}.

Recall that polarizers filter incident light such that light parallel to the polarizer's axis has maximal transmission intensity $T_{max}$ and the light orthogonal to the its axis has minimal transmission intensity $T_{min}$. Up until now, we've assumed that the ratio between $T_{min}$ and $T_{max}$ (known as the extinction ratio) is equal to 0, which is the case for perfect polarizers. Real world polarizers, however, have a non-zero extinction ratio. For instance, elements in the polarization array of our camera have an extinction ratio $\rho$ around 1:500.

In Figure~\ref{fig:light-leakage}, we experimentally demonstrate the non-zero extinction ratio of real world polarizers by setting the angle $\theta$ between two polarizers to $\frac{\pi}{2}$. According to Malus' law, all light should be blocked and the recorded pixel intensity values should be zero. Instead, we visibly see white pixels in the image around the bright tungsten filament in the light bulb.

To account for imperfect polarizers, we revisit Equation~\eqref{eq:closed-form} and re-express the intensity $I$ of light on the sensor in terms of the extinction ratio $\rho$. 

Irradiance $I_0$ incident on the camera can be described by the Jones vector $\vec{E}_0$ where
\begin{align}
    I_0 &= |\vec{E}_0|^2.
\end{align}

This vector expresses the polarization state of light as 
\begin{equation}
    \vec{E} = \begin{bmatrix}
    A_x e^{i\delta_x} \\ A_y e^{i\delta_y}
    \end{bmatrix}
    = \begin{bmatrix}
    X\\Y
    \end{bmatrix},
\end{equation}
where $A_x, A_y$ denote amplitude and $\delta_x, \delta_y$ denote phase delay along axes $x, y$. For convenience, we define axes $x, y$ such that they are aligned with the transmission and extinction axes of the first polarizer $P_1$ respectively.

The extinction ratios $\rho_1, \rho_2$ of polarizers $P_1, P_2$ share the following relationship with amplitudes $A_{1,x}, A_{1,y}$ of waves exiting $P_1$ and amplitudes $A_{2,x}, A_{2,y}$ of waves exiting $P_2$:

\begin{align}
    \rho_1 &= \sqrt{\frac{A_{1,y}}{A_{1,x}}} \\
    \rho_2 &= \sqrt{\frac{A_{2,y}}{A_{2,x}}}.
\end{align}
For simplicity, we assume that both polarizers have the same extinction ratio $\rho_1 = \rho_2 = \rho$, but our derivation also applies to non-uniform extinction ratios.

The first polarizer $P_1$ attenuates the amplitude $A_{y,0}$ of incident light $\vec{E}_0$ along the $y$ axis by a factor 

\begin{align}
    \alpha = \sqrt{\frac{\rho}{\rho+1}}
\end{align}
Assuming that $T_{max}=1$, the light $\vec{E}_1$ exiting $P_1$ subsequently becomes
\begin{equation}
        \vec{E}_1 = \begin{bmatrix}
        1\\
        \alpha
    \end{bmatrix}\odot\vec{E}_0.
\end{equation}

The second polarizer $P_2$ also attenuates light by a factor $\alpha$. However, this attenuation happens along $P_2$'s axes which are not necessarily aligned with $x,y$. By definition, the two axes differ from one another with angle $\theta$. So, we project light $\vec{E}_1$ after $P_1$ onto $P_2$'s axes $u,v$ before applying the attenuation factor $\alpha$.
\begin{equation}
    proj_{\textbf{u,v}}{\vec{E}_2} =
    \begin{bmatrix}
        1 \\
        \alpha
    \end{bmatrix}
    \begin{bmatrix}
        \text{cos}\theta & \text{sin}\theta\\
        -\text{sin}\theta & \text{cos}\theta
    \end{bmatrix}
    \vec{E}_1\label{eq:alpha}
\end{equation}

Finally, we reproject the attenuated vector onto $x,y$ and obtain an expression for the light $\vec{E}_2$ that hits the sensor
\begin{equation}
    \vec{E}_2 = \begin{bmatrix}
        \text{cos}\theta & -\text{sin}\theta\\
        \text{sin}\theta & \text{cos}\theta
    \end{bmatrix} proj_{\textbf{u,v}}{\vec{E}_2}.
\end{equation}

Expanding terms, we derive the following expression of light observed by the sensor $\vec{E}_2$ with respect to $\alpha$ and incident light $\vec{E}_0=\begin{bmatrix}
    X_0\\Y_0
\end{bmatrix}$
\begin{equation}
    \vec{E}_2 = \begin{bmatrix}
        X_0\text{cos}^2\theta+\alpha X_0\text{sin}^2\theta+(\alpha+\alpha^2)(Y_0\text{sin}{\theta}\text{cos}{\theta}) \\
        \alpha^2 Y_0\text{cos}^2\theta+\alpha Y_0\text{sin}^2\theta+(1-\alpha)(X_0\text{sin}{\theta}\text{cos}{\theta}) \\
    \end{bmatrix}.
\end{equation}

Intensity measured by the camera is simply the squared amplitude of the wave $\vec{E}_2$ at the sensor plane, described by
\begin{equation}
    I(\theta, \alpha) = |\vec{E}_2|^2.
\end{equation}

Removing our prior assumption of perfect polarizers, we reformulate Equation~\eqref{eq:closed-form} to take an additional parameter $\alpha$ related to the extinction ratio by Equation~\eqref{eq:alpha}.

\begin{equation}
    \hat{\theta}_1 = \text{arctan}\left(\sqrt{\frac{I(\theta_3, \alpha)}{I(\theta_1,\alpha)}}\right)
\end{equation}

\begin{figure}
    \centering
    \includegraphics[width=\columnwidth]{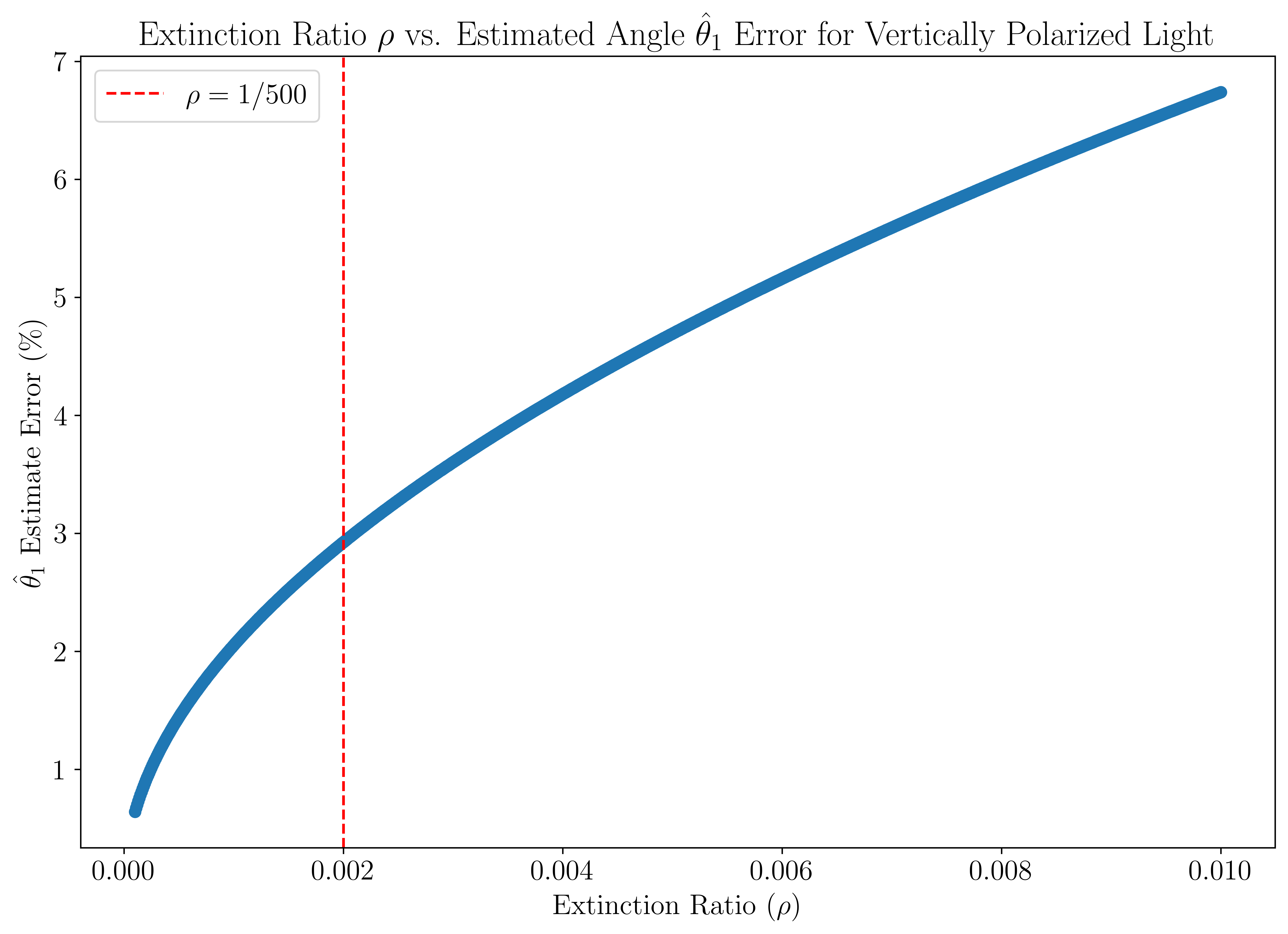}
    \caption{\textbf{Extinction ratio vs. $\hat{\theta}$ estimate error.} As the polarizers' extinction ratio $\rho$ deviates from 0, errors in our estimate of angle $\theta$ grow logarithmically. The extinction ratio of our linear polarizer is drawn in red for reference.}
    \label{fig:extinction-analysis}
\end{figure}

In Figure~\ref{fig:extinction-analysis}, we visualize the effect that imperfect polarizers have on our estimates by plotting estimate error as a function of the extinction ratio $\rho$ (assuming the incident light is vertically polarized). Interestingly, the graph reveals an approximately logarithmic relationship between the error and extinction ratio. For example, on our imaging hardware, which has an extinction ratio of 1:500, our method estimates ${\theta}_1$ with an error of 2.93\%.

\subsection{Highly Polarized Scenes}

Our optical setup measures an HDR scene at a single polarity; this produces measurements distinct from those obtained from an exposure bracket captured by a standard DSLR camera. When dealing with highly polarized scenes (e.g.,~light reflected off the surface of a lake), the angle $\phi$ of the initial linear polarizer $P_1$ noticeably affects the resulting image. To analyze this, we conduct the following experiment:

We position our polarization camera at approximately Brewster's angle relative to the reflective surface of a creek. With the linear polarizer off, we capture an multi-shot exposure bracket and generate 4 HDR images (one image per polarity). Reattaching the linear polarizer, we recapture the scene twice---once with the polarizer nearly vertical ($\phi \approx \pi / 2$) and once with it nearly horizontal ($\phi \approx 0$).

\begin{figure}[t]
    \centering
    \includegraphics[width=\columnwidth]{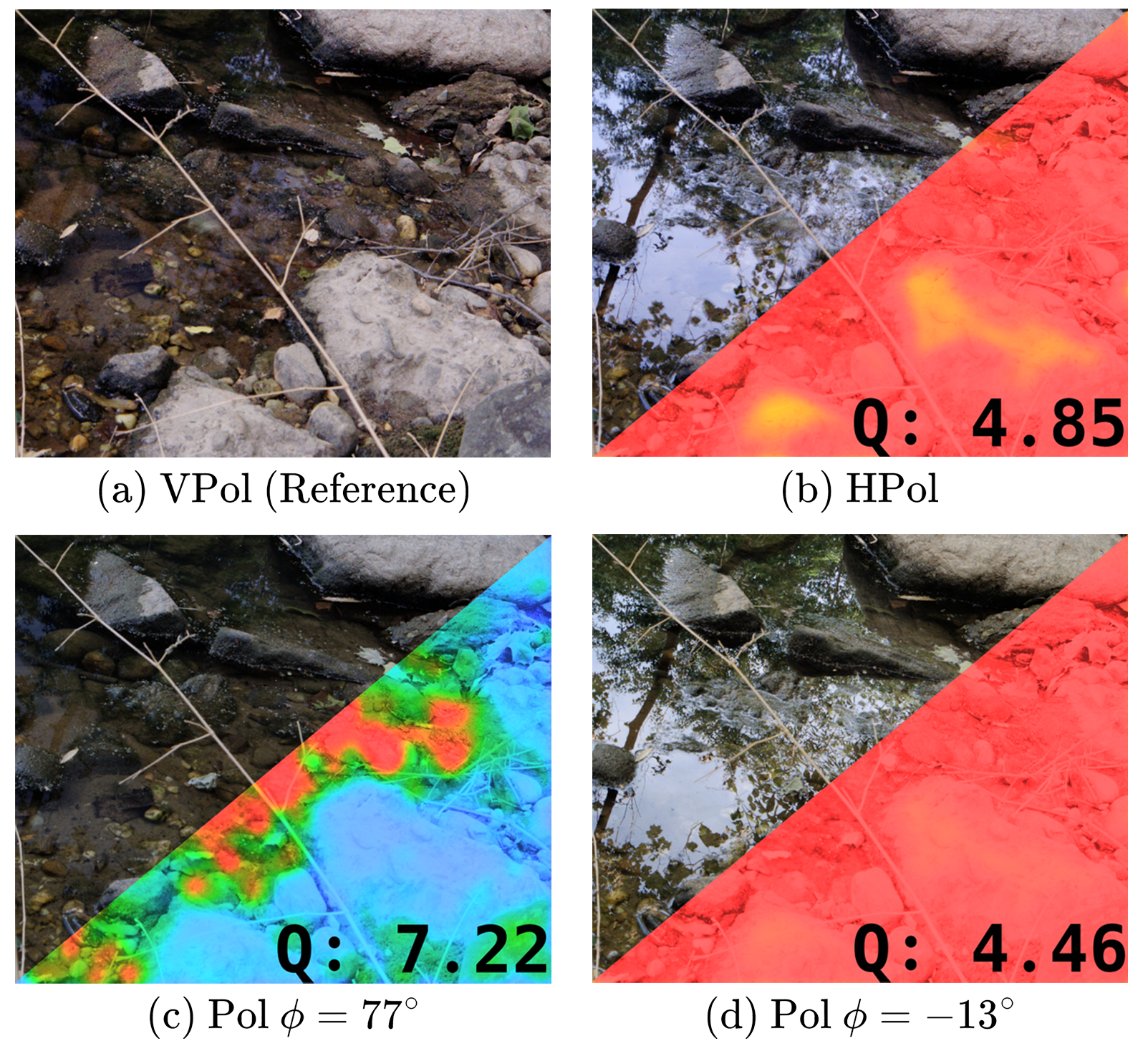}
    \caption{\textbf{Sensitivity to polarized incident light.} Imaging through a vertical polarizer (left column) filters out specular highlights reflecting off of the water's surface. Rotating the polarizer horizontally (right column) allows us to image reflections on the water. Overlays show Q-scores and $P_{det}$ heat maps against the reference image  (a).}
    \label{fig:experiment}
\end{figure}

In Figure~\ref{fig:experiment}, we compare (a) a vertically-polarized, multi-shot HDR reference image against three images: (b) the horizontally-polarized multi-shot HDR image, (c) our method's snapshot HDR image with $\phi=-13^\circ$, and (d) our method's snapshot HDR image with $\phi=77^\circ$. Images captured with the polarizer at a vertical angle filter out specular reflections---allowing us to see through the water's surface---whereas those with the polarizer at a horizontal angle (b, c) do not.

Although our method captures light at a single polarity, we emphasize that this property is not strictly negative. Filtration of light by polarity is useful for numerous applications such as material classification and reflection removal. Consider the example in Figure~\ref{fig:experiment}. Without the use of a polarizer, specular reflections on the creek completely obstruct our view of the rock and sediment beneath the water's surface. Inclusion of a vertical polarizer allows us to filter out these reflections and capture a clear image of the riverbed.

\section{Discussion and Limitations}

Our technique has a few notable limitations, the most significant of which is light efficiency. After passing through the polarizers, only a fraction of the light incident on the camera's initial polarizer makes it to the sensor. Thus our system is best suited for imaging in well-lit environments.

Another limitation associated with our method is its spatial resolution. Similar to mosaiced HDR sensors, our polarization camera trades off spatial resolution to capture multi-exposure information. Advanced demosaicking algorithms could help recover this information~\cite{qiu2021linear}.

\subsection{Mosaiced HDR Sensors}

Mosaiced HDR sensors are powerful tools for single-shot HDR; yet, they are specialized sensors that cost upwards of \$500. By contrast, our proposed system costs as little as \$40 for individuals that own a polarization camera. In addition, our system offers the ability to measure polarity simply by removing the polarizer, which is useful for surface normal estimation, reflection removal, material classification, and more.

\begin{figure*}
    \centering
    \includegraphics[width=\textwidth]{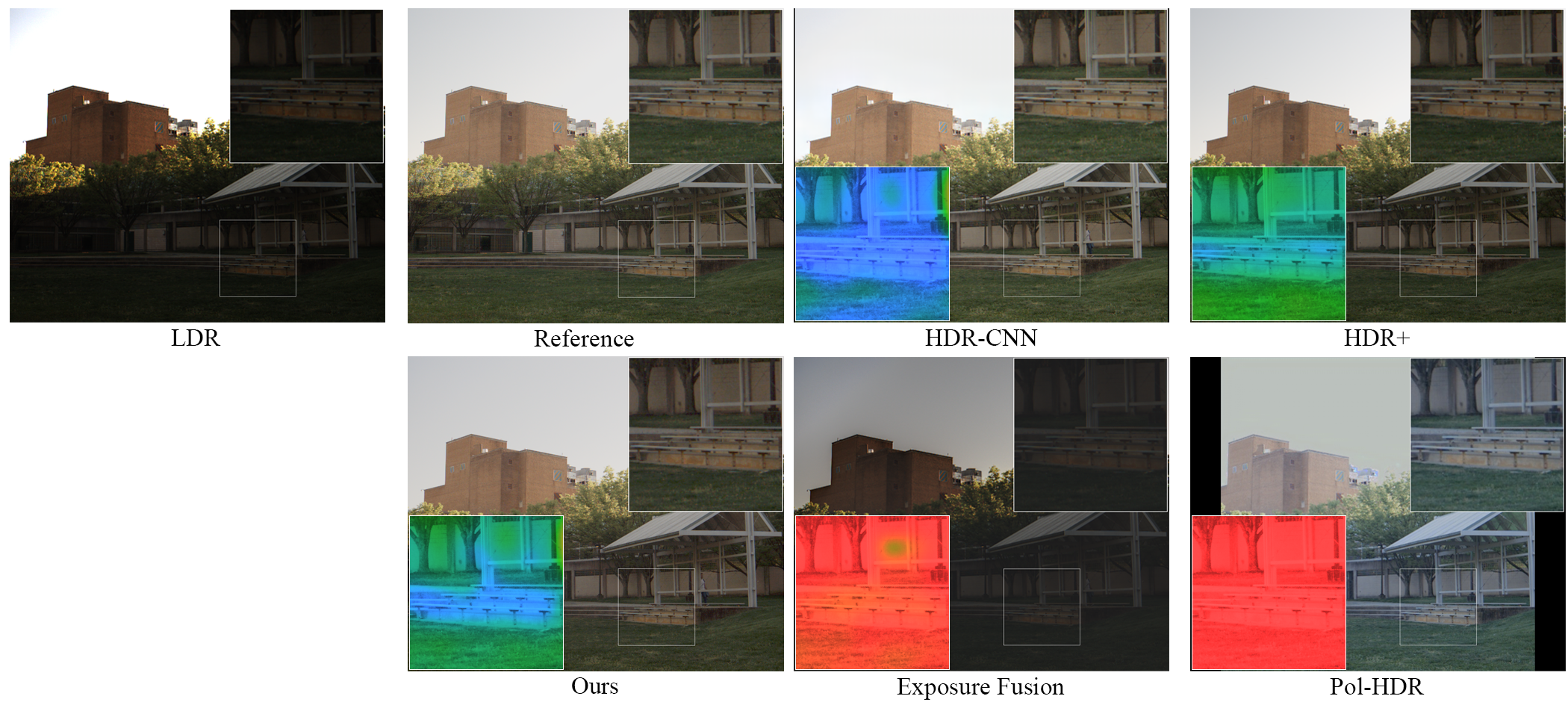}
    \includegraphics[width=\textwidth]{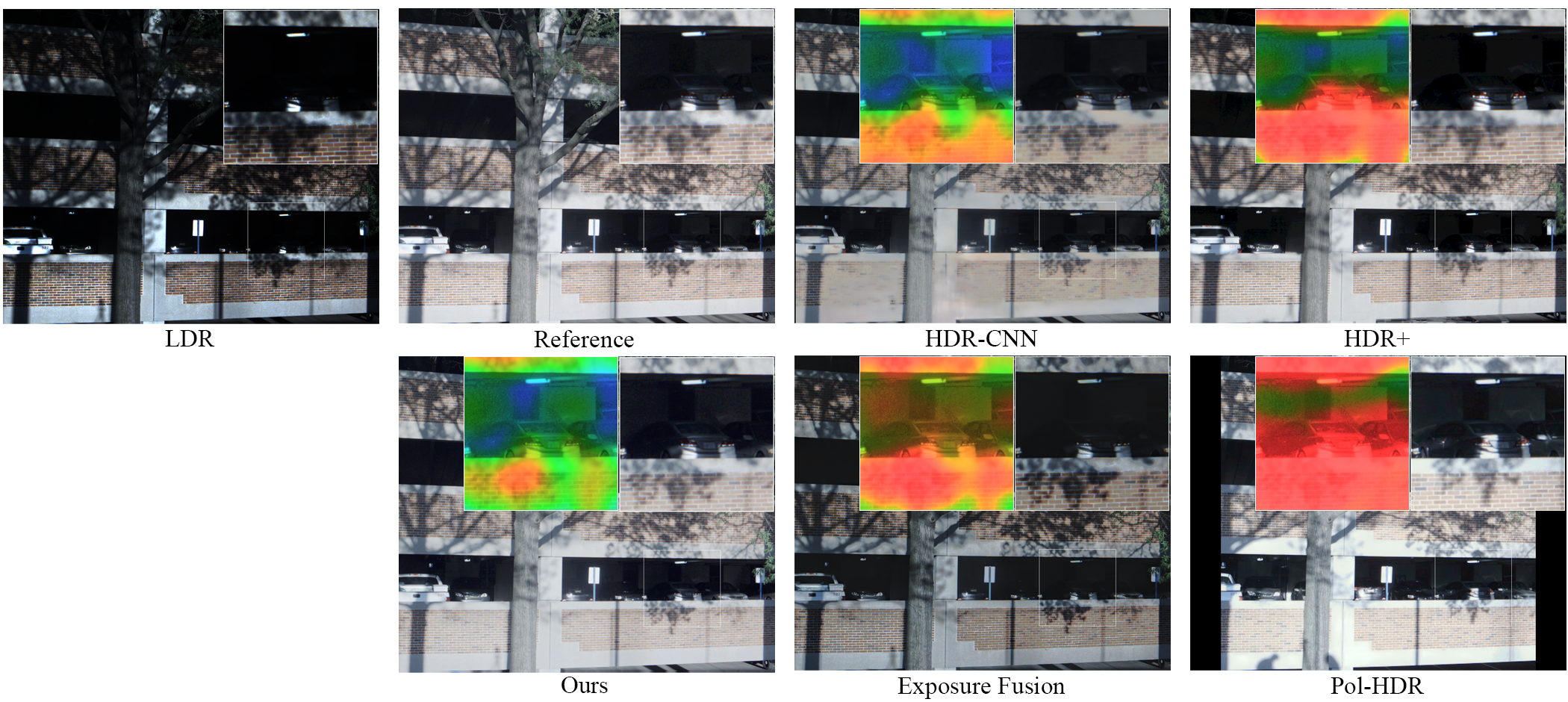}
    \caption{\textbf{Additional results.}}
    \label{additional-results}
\end{figure*}

\section{Conclusion}
In this work, we propose a straightforward and reversible procedure to turn a polarization camera into a snapshot HDR camera. Moreover, we introduce a simple, yet effective, self-calibrating HDR reconstruction algorithm for our system and experimentally demonstrate that our algorithm outperforms existing single-shot and multi-shot HDR reconstruction methods. Finally, we analyze the effects that both imperfect polarizers and highly polarized light have on our results.

Unlike dedicated HDR sensors, our system is dual purpose---the polarizer can be added for HDR imaging or removed for polarization imaging. Our approach thus provides a convenient way for polarization camera users to perform high-quality snapshot HDR imaging at a minimal cost.

\bibliographystyle{IEEEtran}
\bibliography{IEEEabrv,references}

\vfill

\end{document}